# DiPerF: an automated DIstributed PERformance testing Framework

Catalin Dumitrescu* Ioan Raicu* Matei Ripeanu* Ian Foster+*

*Computer Science Department
The University of Chicago
{cldumitr,iraicu,matei }@cs.uchicago.edu

+Mathematics and Computer Science Division
Argonne National Laboratory
foster@cs.uchicago.edu

## Abstract

*We present DiPerF, a distributed performance-testing framework, aimed at simplifying and automating service performance evaluation. DiPerF coordinates a pool of machines that test a target service, collects and aggregates performance metrics, and generates performance statistics. The aggregate data collected provide information on service throughput, on service 'fairness' when serving multiple clients concurrently, and on the impact of network latency on service performance. Furthermore, using this data, it is possible to build predictive models that estimate a service performance given the service load. We have tested DiPerF on 100+ machines on two testbeds, Grid3 and PlanetLab, and explored the performance of job submission services (pre-WS GRAM and WS GRAM) included with Globus Toolkit® 3.2.*

## 1. Introduction

We present DiPerF, a performance-testing framework that aims to simplify and automate service performance evaluation. Multiple threads motivated us to build this framework. Firstly, although performance testing is an 'everyday' task, testing harnesses are often built from scratch for a particular service. DiPerF can be used to test the scalability limits of a service: that is, find the maximum offered load supported by the service while still serving requests with an acceptable quality of service. Secondly, in the context of heterogeneous, geographically distributed clients with different levels of connectivity the actual service performance experienced by clients cannot be gauged based on controlled, LAN-based tests. Therefore significant effort is sometimes required in deploying the testing platform itself. With a wide-area, heterogeneous deployment provided by the PlanetLab [1, 2] and/or Grid3 [3] testbed, DiPerF can provide accurate estimation of the service performance as experienced by such clients. To summarize, DiPerF goal is to become a practical tool for automated service performance evaluation; and, in the long run, to offer service developers the ability to 'outsource' the performance evaluation of their services.

DiPerF coordinates a distributed pool of machines that run clients of a target service, collects and aggregates performance metrics, and generates performance statistics. The data collected provide information on a particular service's maximum throughput, on service 'fairness' when multiple clients access the service concurrently, and on the impact of network latency on service performance from both client and service viewpoint. Using this data it is possible to build empirical performance estimators that link observed service performance (throughput, response time) to offered load. These estimates can be then used as input by a resource scheduler to increase resource utilization while maintaining desired quality of service levels. All steps involved in this process are automated, including dynamic deployment of a service and its clients, testing, data collection, and data analysis.

Automated performance evaluation and result aggregation across a distributed test-bed is complicated by multiple factors. Firstly, the accuracy of the performance metrics collected is heavily dependent on the accuracy of the timing mechanisms used and on accurate time synchronization among the participating machines. DiPerF synchronizes the time between client nodes with a synchronization error smaller than 100ms. Secondly, the scalability of the framework itself is important; otherwise DiPerF will not be able to saturate a target service. We insure scalability by only loosely coupling the participating components. Thirdly, reliability of presented results is important, especially in wide-area environments: we detect client failures during the test that could impact on reported result accuracy. While the results we present here use 89 client machines, we have used DiPerF on a pool of more than 100 machines, and we designed it to scale to 1000s of machines.

We deployed DiPerF on two testbeds, Grid3 and PlanetLab, and tested the performance of job submission services available with Globus Toolkit version 3.2 (GT3.2): the Grid Resource Allocation & Management (pre-WS GRAM) and the Managed Job Factory Service (WS GRAM) [4, 5, 6, 7]. (For the rest of this paper, whenever we refer to pre-WS GRAM, we imply GT3.2 pre-WS GRAM, and whenever we refer to WS GRAM, we imply GT3.2 WS GRAM.)

While our project is still in its infancy, we believe that DiPerF will prove to be a valuable tool for scalability and performance evaluation studies, as well as for automated extraction of service performance characteristics.

The rest of this paper is structured as follows. The following section presents related work. Section 3 describes the DiPerF framework, its deployments, and the services we benchmark in this paper. Measurement results are presented in Section 4. The paper concludes with a brief summary of our experience and future work plans.

## 2. Related Work

Many studies have investigated the performance of individual Grid services. As an example, Zhang et al. [8] compare the performance of three resource selection and monitoring services: the Globus Monitoring and Discovery Service (MDS), the European Data Grid Relational Grid Monitoring Architecture (R-GMA) and Hawkeye. Their experiment uses two sets of machines (one running the service itself and one running clients) in a LAN environment. The setup is manual and each client node simulates 10 users accessing the service. This is exactly the scenario where DiPerF would have proved its usefulness: it would have freed the authors from deploying clients, coordinating them, and collecting performance results, and allow them to focus on optimally configuring and deploying the services to test, and on interpreting performance results.

The Globus Toolkit's job submission service test suite [9] uses multiple threads on a single node to submit an entire workload to the server. However, this approach does not gauge the impact of a wide-area environment, and does not scale well when clients are resource intensive which means that the service will be relatively hard to saturate.

The Network Weather Service (NWS) [10, 11] is a distributed monitoring and forecasting system. A distributed set of performance sensors feed forecasting modules. There are important differences to DiPerF. First, NWS does not attempt to control the offered load on the target service but merely to monitor it. Second, the performance testing framework deployed by DiPerF is built on the fly, and removed as soon as the test ends; while NWS sensors aim to monitor services over long periods of time. Similarly, NETI@home [12], Gloperf [13], and NIMI [14] focus on monitoring service or network level performance.

NetLogger [15] targets instrumentation of Grid middleware and applications, and attempts to control and adapt the amount of instrumentation data produced in order not to generate too much monitoring data. NetLogger is focusing on monitoring, and requires code modification in the clients; furthermore, it does not address automated client distribution or automatic data analysis. Similarly, the CoSMoS system [16] is geared toward generic network applications.

GridBench [17] provides benchmarks for characterizing Grid resources and a framework for running these benchmarks and for collecting, archiving, and publishing results. While DiPerF focuses on performance exploration for entire services, GridBench uses synthetic benchmarks and aims to test specific functionalities of a Grid node. However, the results of these benchmarks alone are probably insufficient to infer the performance of a particular service.

Finally, Web server performance has been a topic of much research. The Wide Area Web Measurement (WAWM) Project for example designs an infrastructure distributed across the Internet allowing simultaneous measurement of web client performance, network performance, and web server performance [18]. Banga et al. [19] measure the capacity of web servers under realistic loads. Both systems could have benefited from a generic framework such as DiPerF.

## 3. The DiPerF Framework

DiPerF consists of two major components: the *controller* and the *testers* (Figure 1). A user of the framework provides to the controller the address of the target service to be evaluated and the client code for the service. The controller starts and coordinates a performance evaluation experiment: it receives the client code, distributes it to testers, coordinates their activity, and collects and aggregates their performance measurements. Each tester runs the client code on its machine, and times the (RPC-like) network calls this code makes to the target service. Finally, the controller collects all the measurements from the testers and performs additional operations (e.g., reconciling time stamps from various testers) to compute aggregated performance views.

The interface between the tester and the client code can be defined in a number of ways (e.g., by using library calls); we take what we believe is the most generic avenue: clients are full blown executables that make one RPC-like call to the service.

The framework is supplied with a set of candidate nodes for client placement, and selects those available as testers.. In future work, we will extend the framework to select a subset of available tester nodes to satisfy specific requirements in terms of link bandwidth, latency, compute power, available memory, and/or processor load. In its current version, DiPerF assumes that the target service is already deployed and running.

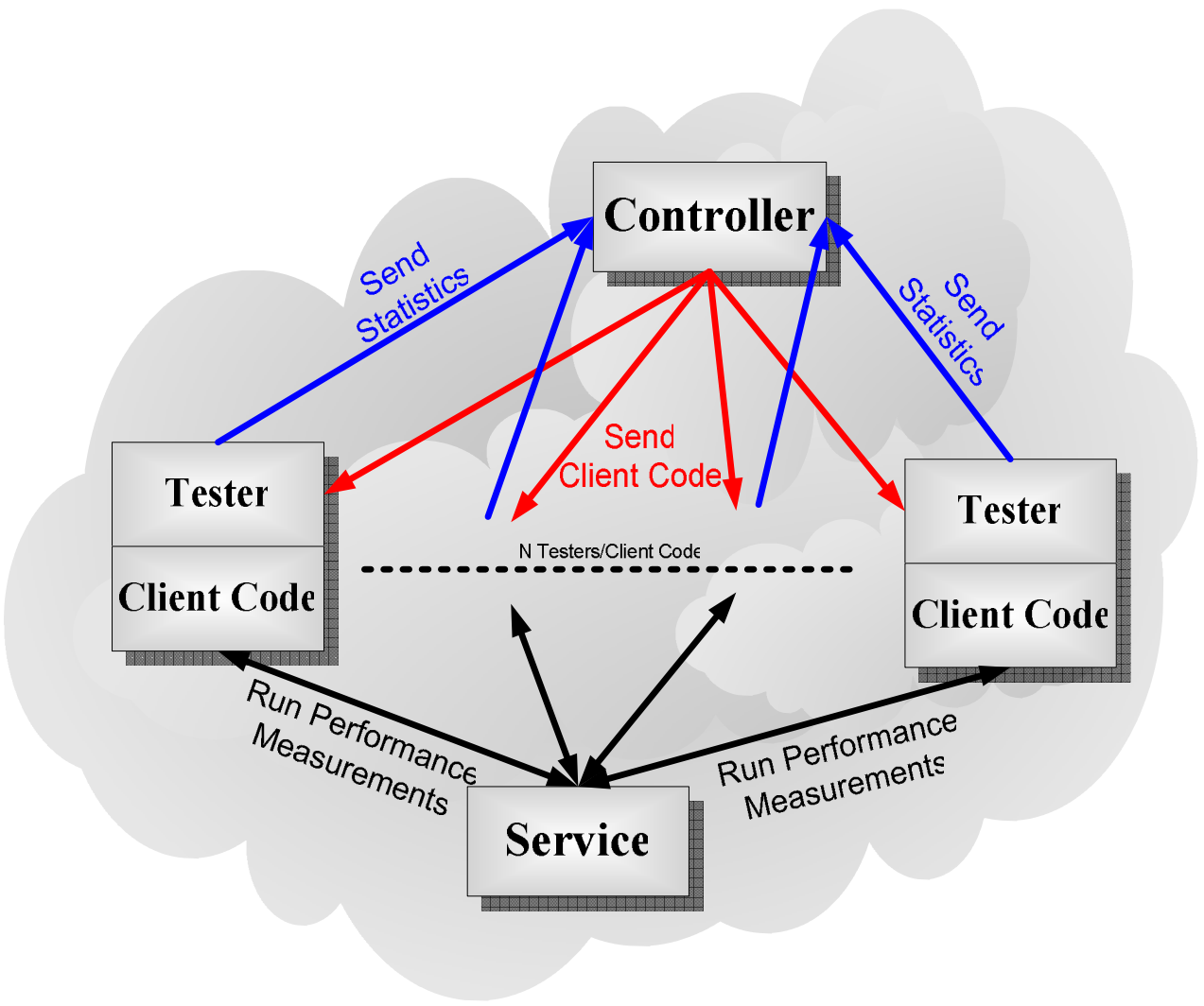

**Figure 1: DiPerF framework overview**

Some metrics are collected directly by the testers (e.g., response time), while others are computed at the controller (e.g., throughput and service fairness). Additional metrics (e.g. network related metrics such as network throughput, size of data transmitted, etc), measured by clients can be reported, through an additional interface, to the testers and eventually back to controller for statistical analysis. Testers send performance data to controller while the test is progressing, this the service evolution can be visualized 'on-line'.

Communication among DiPerF components is ssh based. When a client fails, we rely on the underlying protocols (i.e. whatever the client uses such as TCP, UDP, HTTP, pre-WS GRAM, etc) to signal an error which is captured by the tester which is in turn sent to the controller to delete the client from the list of the performance metric reporters. A client could fail because of various reasons: 1) predefined timeout which the tester enforces, 2) client fails to start (i.e. out of memory - OS client machine related), 3) and service denied or service not found (service machine related). Once the tester is disconnected from the controller, it stops the testing process to avoid loading the target service with requests which will not be aggregated in the final results.

## 3.1 Technical Issues

We summarize below the technical issues we addressed as we developed the framework: client code distribution, time synchronization across the testbed, and performance metrics aggregation.

**3.1.1 Client Code Distribution:** The mechanisms used to distribute client code (e.g., scp, gsi-scp, or gass-server) vary with the deployement environment. Since ssh-family utilities are deployed on just about any Linux/Unix, we base our distribution system on scp-like tools.

**3.1.2 Clock Synchronization:** We rely heavily on time synchronization when aggregating results at the controller. Thus, we have to insure all clients' times are synchronized. Synchronization need not be performed on-line; instead, we can compute the offset between local and global time and apply that offset when analyzing aggregated metrics.

Several options are available to synchronize the time between machines; one example is NTP [20], which some PlanetLab nodes use to synchronize their time. In a previous study [20], hosts had a mean delay of 33ms, median 32ms, and a standard deviation 115ms from their peer hosts used to synchronize their time with NTP.

At the time that performed our experiments, we found that most of the nodes in our testbed on PlanetLab were not very well synchronized, with some nodes having synchronization differences in the thousands of seconds. Therefore our framework assumes the worst: no clock synchronization mechanism is provided by the deployment platform. To ensure that a mechanism exists to synchronize time among all nodes within tens of milliseconds accuracy, we implement a timer component that allows all other nodes to query for 'global' time.

We handle time synchronization with a centralized time-stamp server that allows time mapping to a common base. The time-stamp server is lightweight and can easily handle the 100+ clients used in our testing. Currently each client synchronize its clock every five minutes, so based on the observed time-stamp server load, we believe this scheme should handle easily the framework's target size of thousands of clients.

We have measured the latency from over 100 clients (deployed on Planetlab machines) to our timestamp server over a period of almost 2 hours. During this interval that the (per-node) latency in the network remained fairly constant and the majority of the clients had a network latency of less than 80ms. The accuracy of the synchronization mechanism we implemented is directly correlated with the network latency and its variance, and in the worst case (non-symmetrical network routes), the timer can be off by at most the network latency. Using our custom synchronization component, we observed a mean of 62ms, median 57ms, and a standard deviation 52ms for the time skew between nodes in our PlanetLab testbed.

Given that the response time of the services we are evaluating in this paper is roughly one order of magnitude larger we believe the clock synchronization technique we implement does not distort our results.

**3.1.3 Client Control and Performance Metric Aggregation.** The controller starts each tester with a predefined delay (specified as a command line argument when the controller is started) in order to gradually build up the load on the service (Figure 2).

A tester understands a simple description of the tests it has to perform. The controller sends test descriptions when it starts a tester. The most important description parameters are: the duration of the test experiment, the time interval between two concurrent client invocations, the time interval between two clock synchronizations, and the local command that has to be invoked to run the client. The controller also specifies the addresses of the time synchronization server and the target service.

Individual testers collect service response times. The controller's job is to aggregate these service response times, correlate them with the offered load and with the start/stop time of each tester and infer service throughput, and service 'fairness' among concurrent clients.

Since all metrics collected share a global time-stamp, it becomes simple to combine all metrics in well defined time quanta (seconds, minutes, etc) to obtain an aggregate view of service performance. This data analysis is automated at the user-specified time granularity.

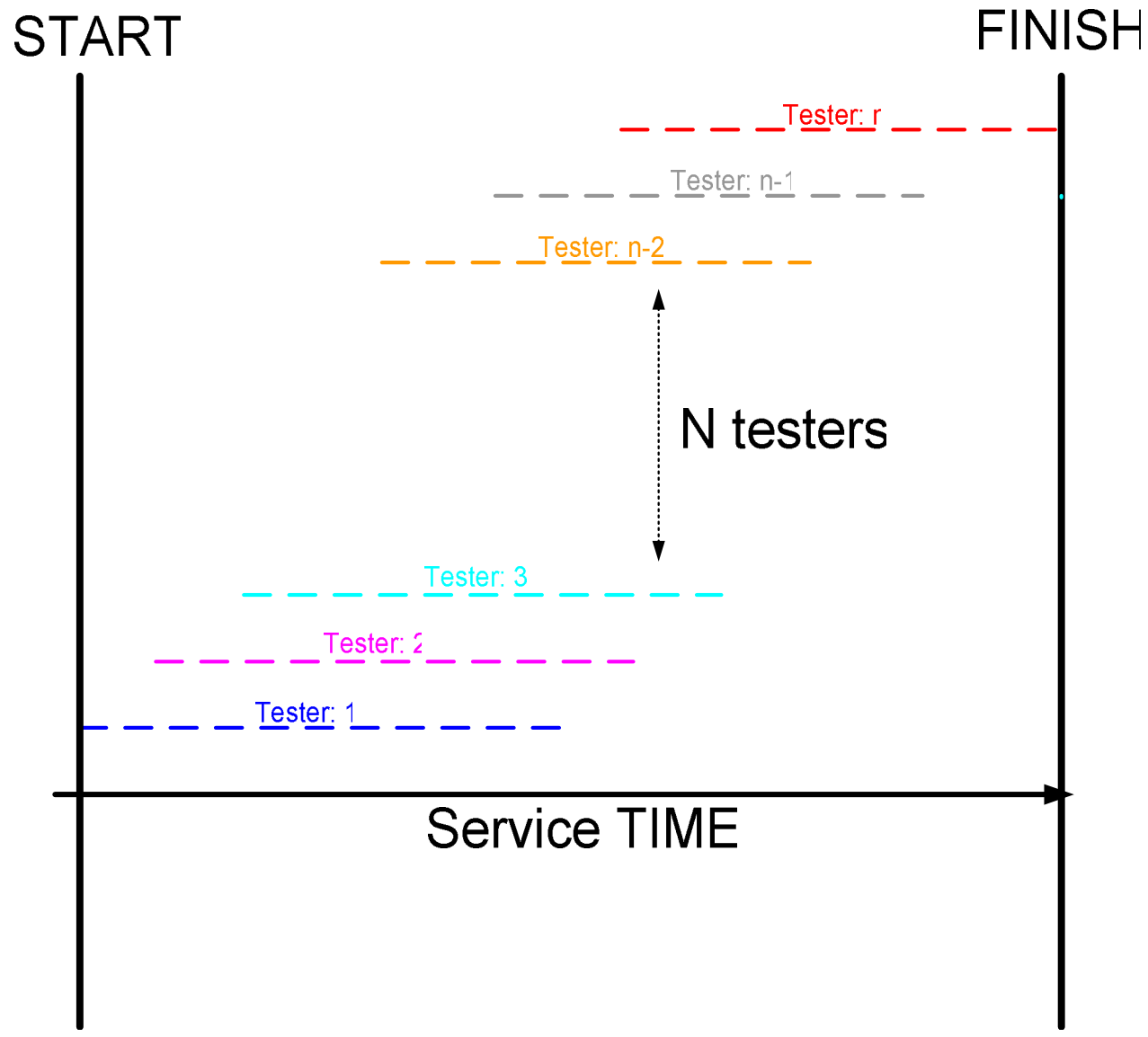

**Figure 2: Aggregate view at the controller. Each tester synchronizes its clock with the time server every five minutes. The figure depicts an aggregate view of the controller of all concurrent testers.**

## 3.2 Target Services

We are evaluating two implementations of a job submission service bundled with various versions of the Globus Toolkit:

- GT3.2 pre-WS GRAM performs the following steps for job submission: a gatekeeper listens for job requests on a specific machine; performs mutual authentication by confirming the user's identity, and proving its identity to the user; starts a job manager process as the local user corresponding to authenticated remote user; then the job manager invokes the appropriate local site resource manager for job execution and maintains a HTTPS channel for information exchange with the remote user.
- GT3.2 WS GRAM, a WS-based job submission service, performs the following steps: a client submits a *createService* request which is received by the Virtual Host Environment Redirector, which then attempts to forward the *createService* call to a User Hosting Environment (UHE) where mutual authentication / authorization can take place; if the UHE is not created, the Launch UHE module is invoked; WS GRAM then creates a new Managed Job Service (MJS); MJS submits the job into a back-end scheduling system [21].

We note that both pre-WS GRAM and WS GRAM are complex services: a job submission, execution, and result retrieval sequence may include multiple message exchanges between the submitting client and the service. In our experiments the submission service creates the jobs on the local machine through the fork interface. As future work, we will be performing similar experiments for GT4.0 WS GRAM. The WS GRAM developers tell us that because the GT4.0 implementation models jobs as lightweight WS-Resources rather than relatively heavyweight Grid services, performance should improve significantly in 4.0 relative to the 3.2 WS GRAM results reported in this paper. We look forward to using our framework to validate that expectation.

# 4. Performance Measurements

The metrics reported by the framework are:

- *service response time* or time to serve a request, that is, the time from when a client issues a request to when the request is completed minus the network latency and minus the execution time of the client code,
- *service throughput*: number of jobs completed successfully by the service averaged over a short time interval,
- *offered load*: number of concurrent service requests (per second),
- *service utilization* (per client): ratio between the number of requests served for a client and the total number of requests served by the service during the time the client was active, and
- *service fairness* (per client): ratio between the number of jobs completed and service utilization.

We ran our experiments with 89 client machines for pre-WS GRAM and 26 machines for WS GRAM, distributed over the PlanetLab testbed and the University of Chicago CS cluster (UofC). We ran the target services on an AMD K7 2.16GHz and the controller on an Intel PIII 600 MHz, both located at UofC. These machines are connected through 100Mbps Ethernet LANs to the Internet and the network traffic our tests generates is far from saturating the network links.

The actual configuration the controller passes to the testers is: testers start at 25s intervals and run for one hour during which they start clients at 1s intervals (or as soon as the last client completed its job if the time the client execution takes more than 1s). The client start interval is a tunable parameter, and is set based on the granularity of the service tested. In our case, since both services (pre-WS GRAM and WS GRAM) quickly rose to service response time of greater than 1s, for the majority of the experiments, testers were starting back-to-back clients. Experiments ran for a total of 5800s and 4200s for pre-WS GRAM and WS GRAM respectively. (The difference in total execution time comes from the different number of testers used). Testers synchronize their time every five minutes. The time-stamp server is another UofC computer.

For pre-WS GRAM, the tester input is a standalone executable that was run directly by the tester, while for the WS pre-WS GRAM, the input is a jar file and we assume that Java is installed on all testing machines in our testbed.

In the figures below, each series of points representing a particular metric is also approximated using a polynomial and/or a moving average in order to show the trend for the metric over the entire period. The polynomial approximations have been computed for all the data in all experiments, and can be used to build empirical models to predict real service behavior. Such models would be useful to a dynamic resource scheduler that seeks to achieve maximum utilization of resources while maintaining a certain QoS.

## 4.1. GT3.2 pre-WS GRAM

Figures 3-5 show results from our pre-WS GRAM job submission experiments. At peak, between seconds 2400 and 3500, all 89 testers run concurrently.

Service response time starts out at about 700ms per job, and increases relatively slowly as offered load increases. As the service load surpassed about 33 clients up to 89 concurrent clients (825s to 2225s in the experiment), the service response time, which is already about 7s per job, starts to fluctuate significantly and increase at a faster rate than for the first 33 machines (0s to 825s of the experiment). A similar phenomenon occurs at the end of the experiments when again at a load of about 33 machines, the service response time stabilizes and starts decreasing uniformly. The dashed line depicts a polynomial approximation of the service response time. The solid line depicts a moving average (over a 160 second window) of the response times, which provides a good approximation under normal load (less than 33 concurrent machines) but an increasingly rough approximation as the load approaches the maximum number of nodes of 89.

We conclude that service capacity (or maximum service throughput) for this service deployment is reached with around 33 concurrent clients. The service could probably handle more clients but this would only result in increased response time. It is interesting to note that jobs in sequential processing take about 700ms to run while after running the experiment over the 89 machines over 5780 seconds, 8025 jobs were completed successfully, which nets 720ms per job. The fact that the average time to complete a job remains virtually unchanged when multiple clients concurrently access the service is evidence that each job uses the full capacity of the resources at the service, and hence the service's resource utilization does not increase or decrease with an increasing service load.

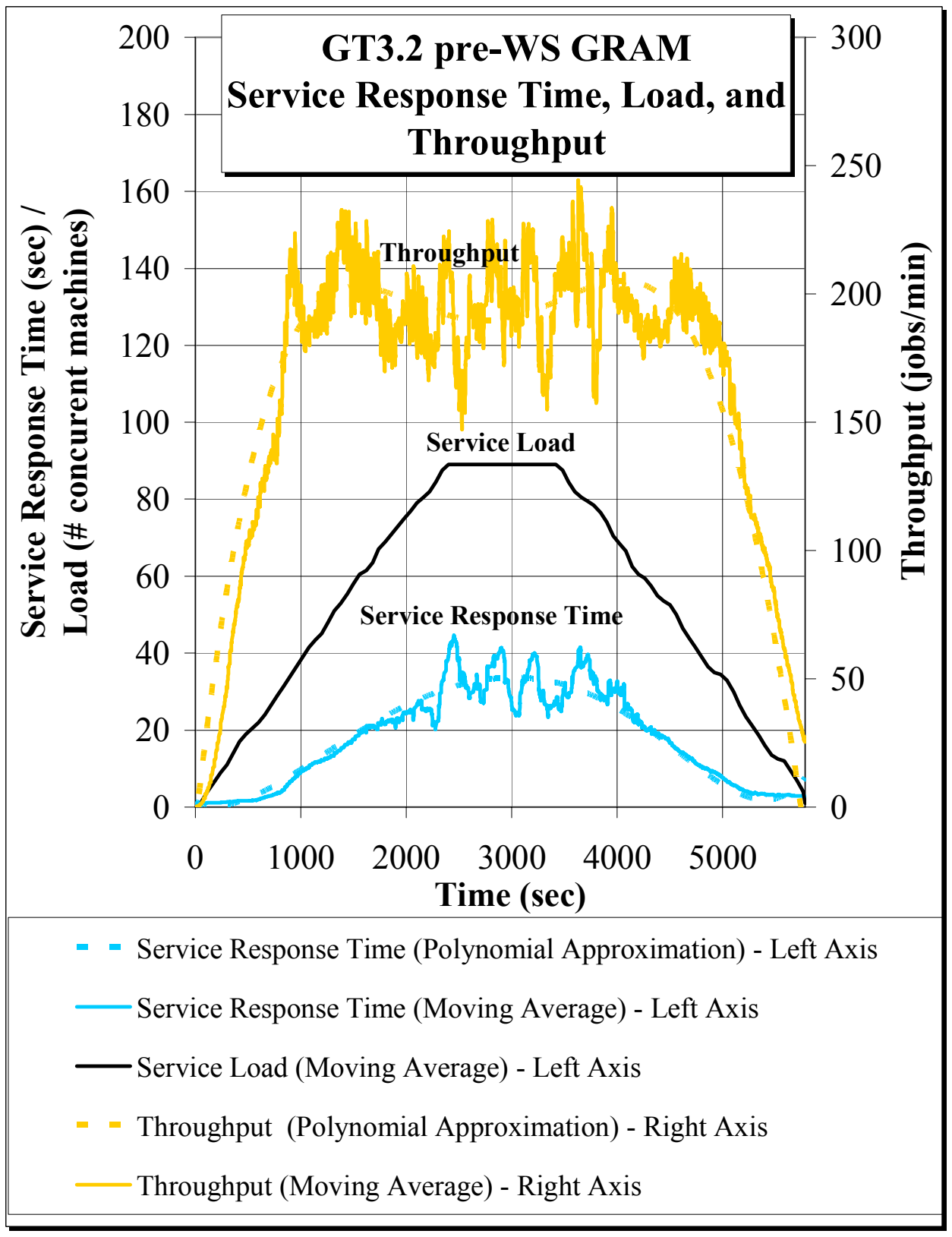

**Figure 3: GT3.2 pre-WS GRAM: Service response time, throughput, and service load**

Despite a wide range of network latencies observed, service performance remains largely unaffected, which indicates that the time to serve a request is not dominated network latencies. We estimate that the service performance is CPU bound (CPU usage larger than 90% when serving sequential requests). To ensure that creating and destroying the large number of processes the submitted workload involved, was not loading the CPU to unacceptable levels, we verified that we could run the same workload locally (not through the GRAM job submission mechanism) in about 200 seconds. The same workload via job submission took about 5780 seconds, which shows that the actual job execution consumed relatively few resources and the job submission mechanism (GT3.2 pre-WS GRAM) actually consumed the majority of the resources.

Service throughput (measured as the number of requests that complete in a given minute) is the second metric we report. We reiterate that the dashed solid line is a polynomial approximation of the throughput, while the solid line is a moving average approximation. We note that throughput measurements also indicate that throughput peaks at about 33 clients, which supports the claim that the service capacity is reached at 33 concurrent clients. Additionally, increased variability can be observed in the polynomial approximation.

Unlike Figure 3, Figures 4 and 5 present machine (tester) IDs on the *x*-axis. Each machine is assigned a unique ID ranging from 1 to the number of machines used in the experiment; IDs are assigned based on machine relative start time: machine with ID 1 starts first and machine with ID 89 starts last. The values presented on the *y*-axis present a given metric computed for a particular client over the duration of the experiment in which the load was at its peak and all clients were concurrently running; in this experiment, this represents about 1100 seconds. For example, in Figure 4 presents the total service utilization and service fairness per client.

The solid line presents *service fairness*, the ratio between the number of jobs completed and the total service utilization. We see that the service gives a relatively equal share of resources to the clients. Figure 5 also supports the claim that the service gave an equal share of resources to the clients: here, each bubble's surface area denotes the number of jobs completed, which monotonically decreases as the load increases, and monotonically increase as the load decreases. Note that the first few machines (as well as the last few machines) have a lower average aggregate load, which means that they had less competition for the resources, and hence had more jobs completed.

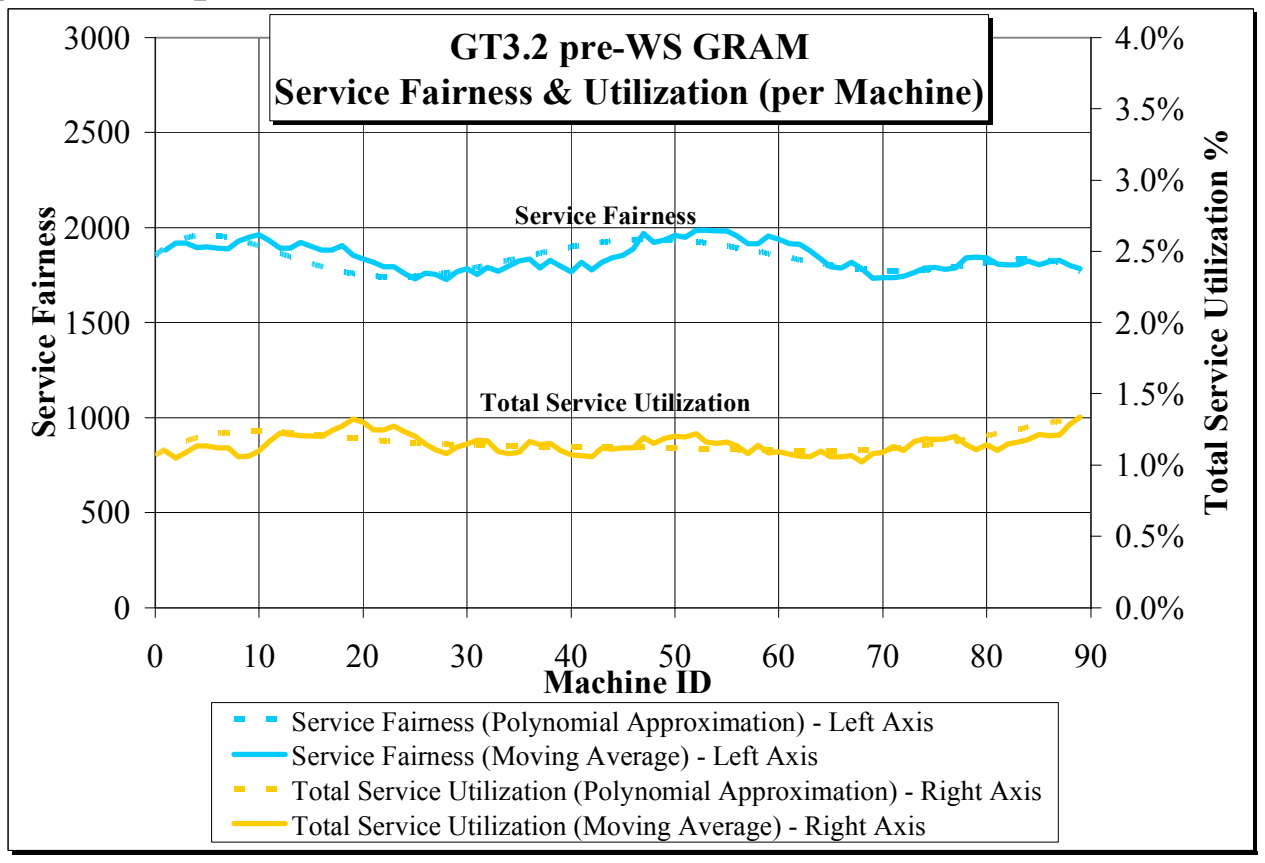

**Figure 4: GT3.2 pre-WS GRAM Service utilization per machine**

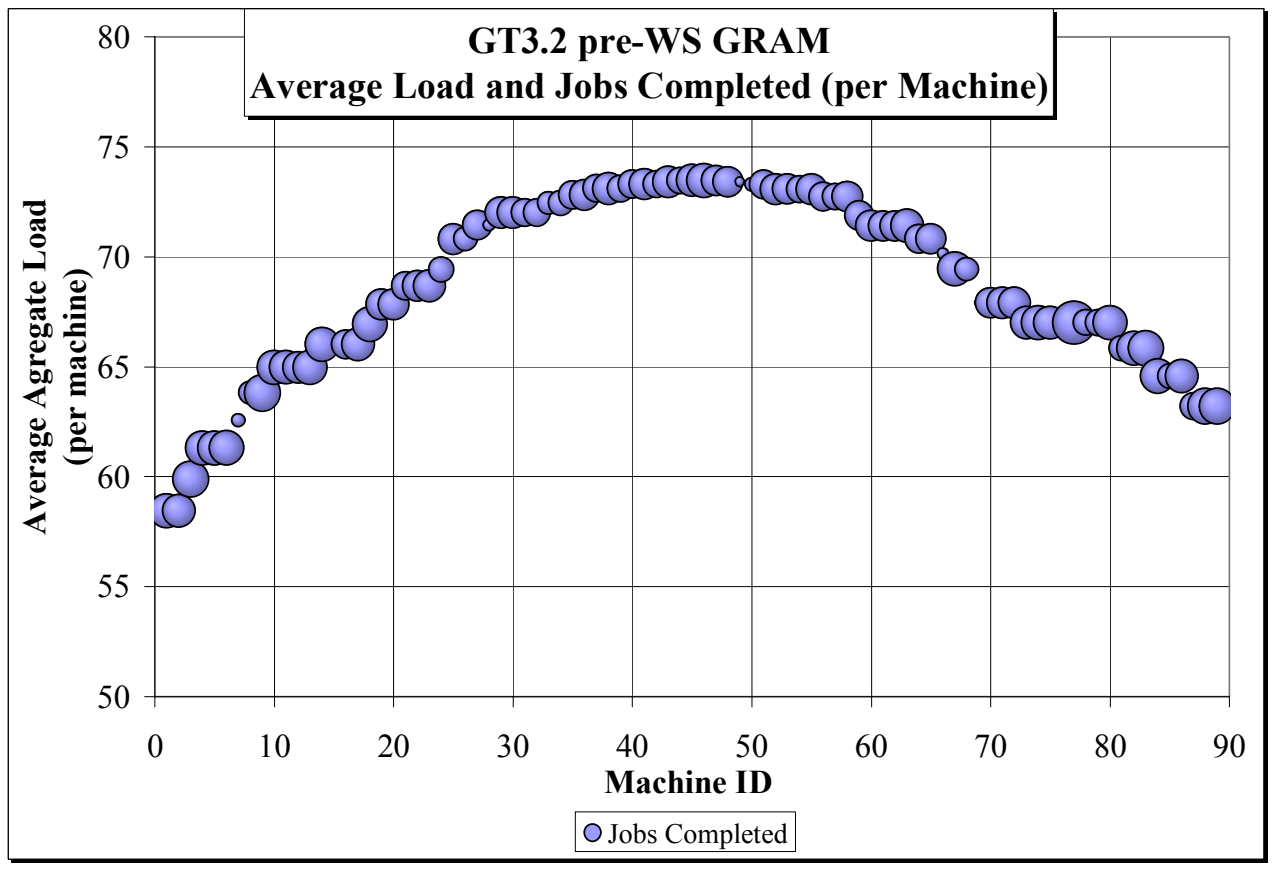

**Figure 5: GT3.2 pre-WS GRAM: Average aggregate load vs. number of requests completed per machine. The (x,y) coordinate of the bubble are the machine ID the average aggregate service load respectively; the size of each bubble is proportional to the number of jobs completed by each client**

## 4.2 GT3.2 WS GRAM

The next set of experiments presents job submission performance for GT3.2 WS GRAM.

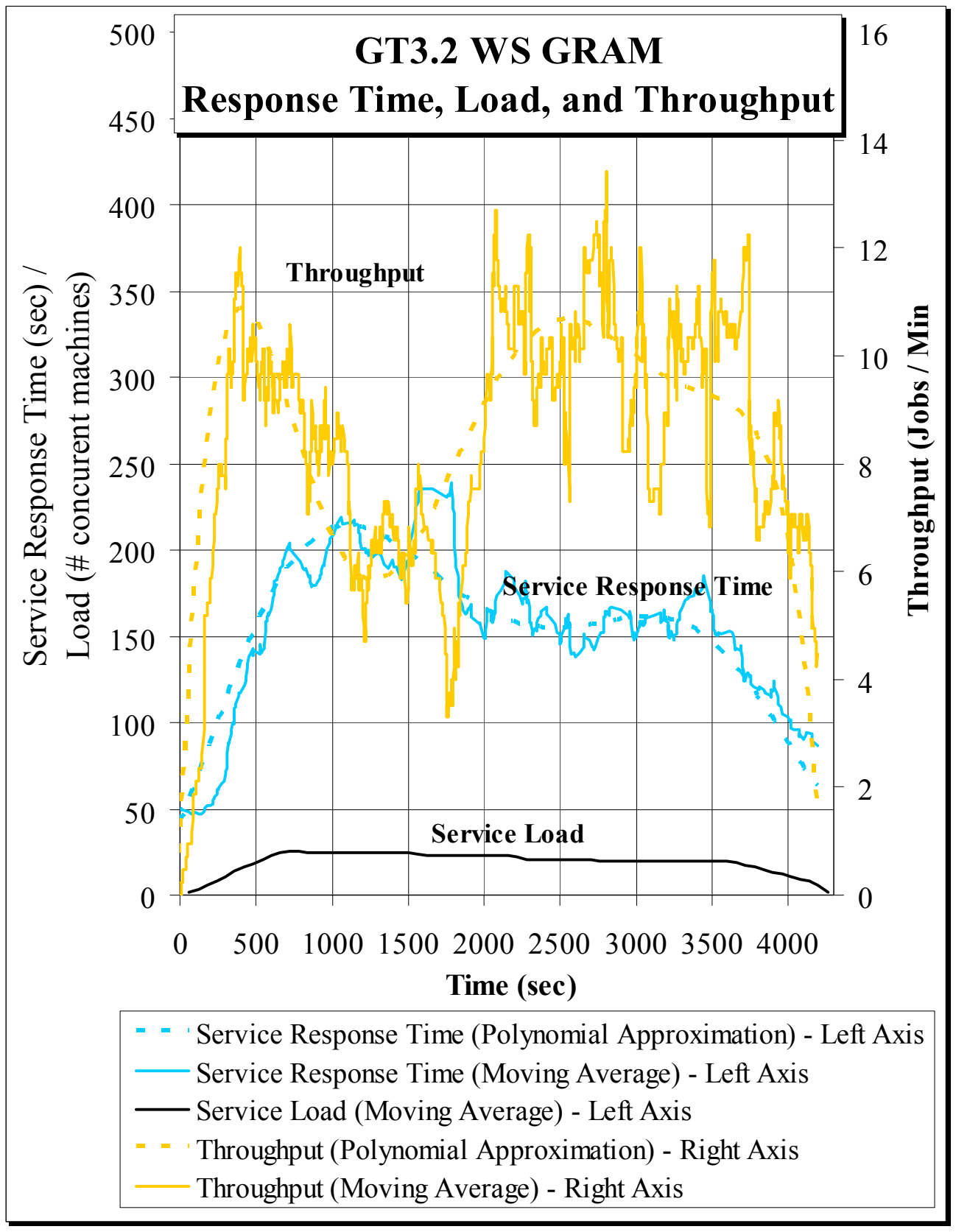

**Figure 6: GT3.2 WS GRAM: Response time, throughput, and load**

We initially attempted to use the same number of clients (89) as for the pre-WS GRAM tests, but the service did not fail gracefully: after serving a small number of requests, the service stalled and all clients attempting to access the service failed. We ultimately used a pool of 26 machines in testing WS GRAM.

The results of Figure 6 show service capacity peaks at around 20 concurrent machines. This result is supported by the fact that the throughput flattens out around the time that 20 machines are accessing the service in parallel, but when the number of machines increases to 26, the throughout decreases dramatically, until a few clients start failing. Once the number of machines decreases to 20 (due to failed clients), the throughout returns to about 10 jobs per minute. Note the service response time decreases as well, once the number of machines stabilized at the service's capacity.

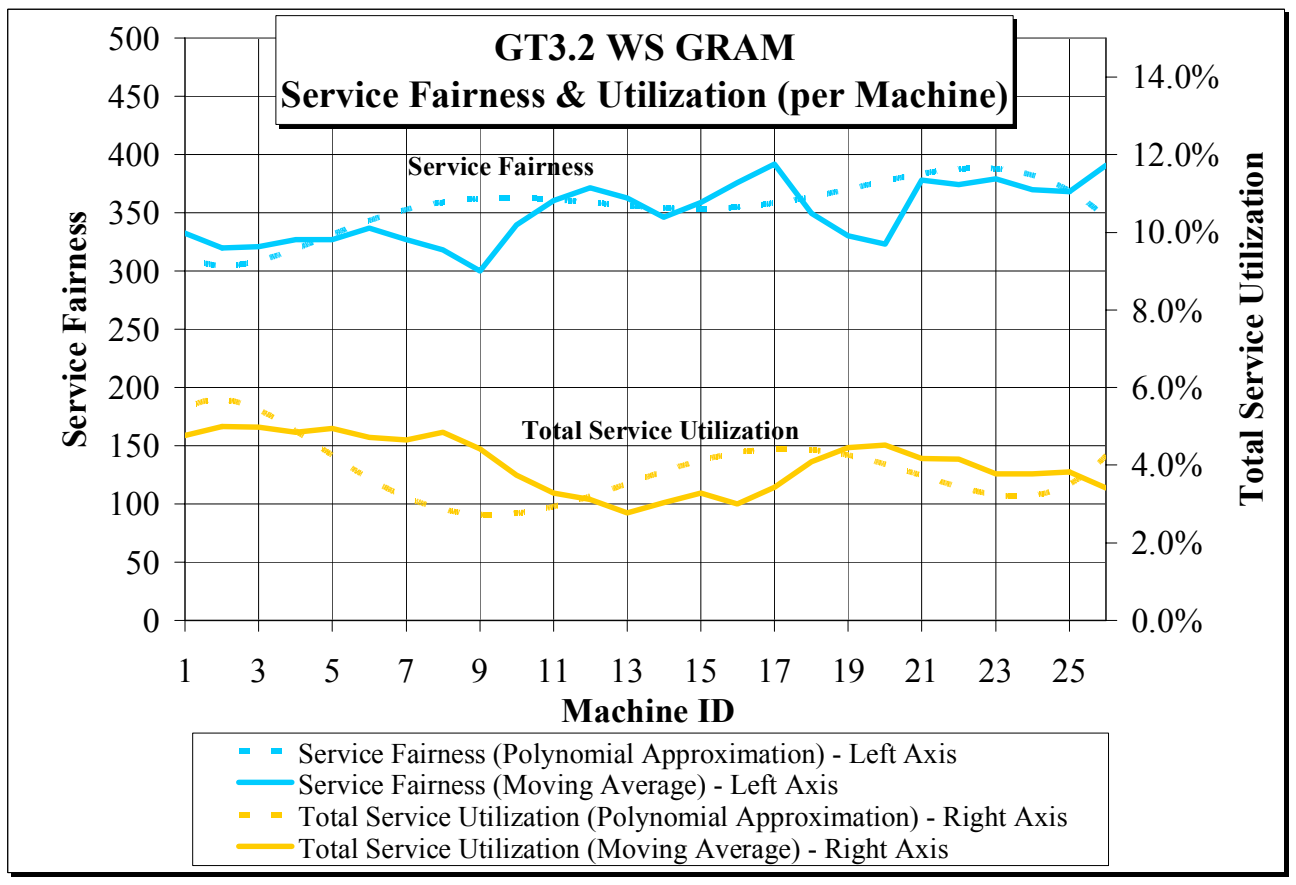

**Figure 7: GT3.2 WS GRAM: Service utilization per machine**

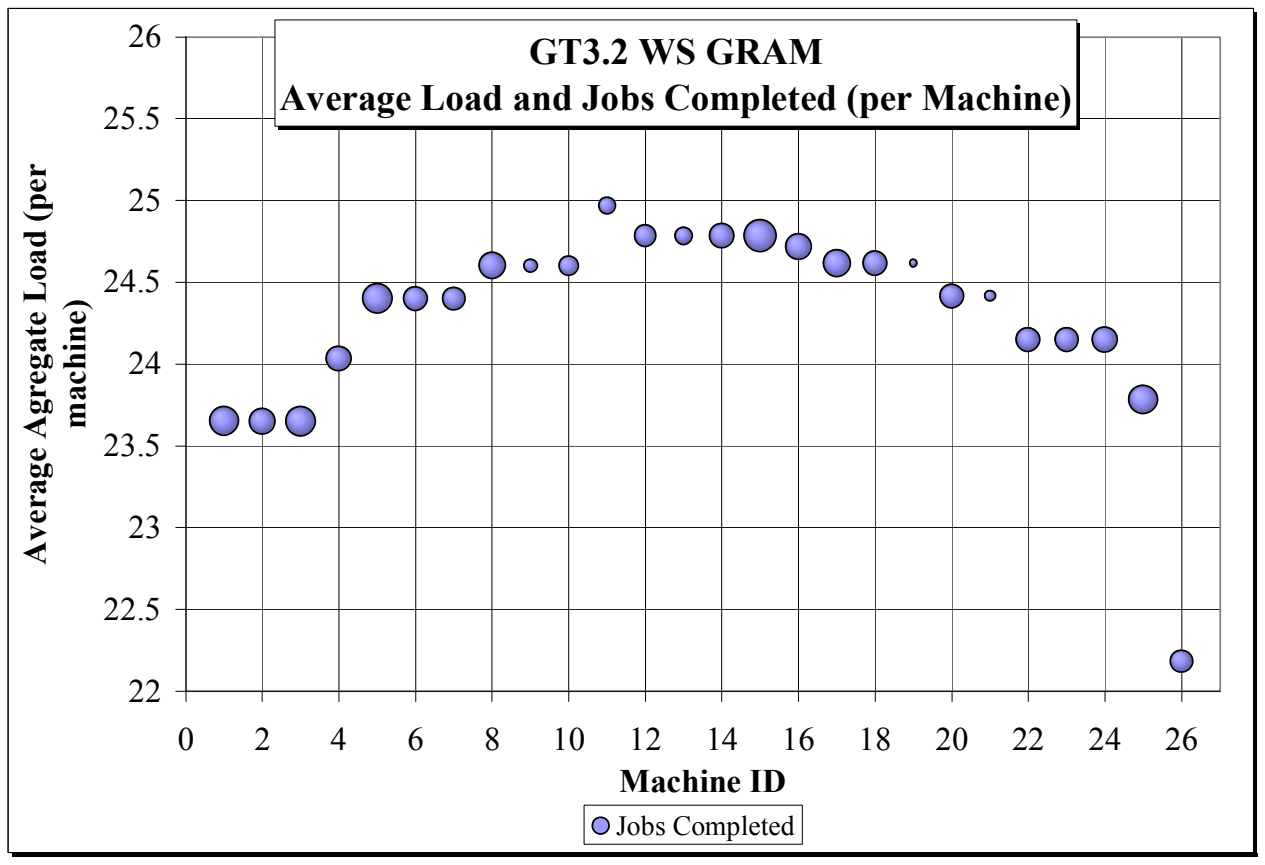

**Figure 8: GT3.2 WS GRAM: Average aggregate load and the number of jobs completed per machine; see Figure 5 for the (x,y) coordinates**

Figure 7 shows that the service fairness varies significantly more than it did for pre-WS GRAM, however based on Figure 8, it seems that only a few clients are not given equal share, which is evident from the few bubbles that have a significantly smaller surface area.

### 4.3 HTTP

Both pre-WS GRAM and WS GRAM job submission services take hundreds of milliseconds or more to complete a single request (even minutes for a heavily loaded instance). In order to show that DiPerF also works for finer granularity services while maintaining accurate results we tested an Apache HTTP server and used "wget" (as the client code) to invoke a CGI script over HTTP. We used the default HTTP server configuration, and 125 PlanetLab machines to perform this experiment, with each client starting at an offset of 25 seconds and executing at most 3 jobs per second per client. We found that the 125 clients were able to saturate the HTTP service and both the throughput and service response time yielded consistent results.

## 5. Summary and Future Work

We have presented early experiences with DiPerF, a distributed performance-testing framework designed to simplify and automate service performance evaluation. DiPerF coordinates a pool of machines that test a target service, collects and aggregates performance metrics, and generates performance statistics. The aggregate data collected provide information on service throughput, on service 'fairness' when serving multiple clients concurrently, and on the impact of network latency on service performance. Furthermore, using the collected data, it is possible to build predictive models that estimate service performance as a function of service load.

We have applied DiPerf to two GT3.2 job submission services, pre-WS GRAM and WS GRAM, and found peak throughput of about 200 and 10 requests per minute, respectively. Average service response time under 'normal' load was about 700ms and 50s respectively. Average service response time under 'heavy' load was about 35s and 150s respectively. We also observed that under heavy load the pre-WS GRAM service allocates resources more evenly among clients when compared to the WS GRAM service.

We plan to perform similar experiments in the near future for GT4.0 pre-WS GRAM. GT4.0 developers assert that GT4.0 usage of lightweight WS-Resources improves performance significantly compared to GT3.2 WS GRAM performance reported here. We plan to perform tests for other services (of various granularities) and in parallel, improve our framework for collecting other metrics required in WAN environments. We also plan to investigate the accuracy of the empirical models we can build using DiPerF, validate them, and find suitable applications where they can be used to improve resource allocation decisions. Finally, we plan to investigate the scalability of DiPerF and its various components (controller, tester, communication protocols, etc) in order to back up the claims we made in this paper that DiPerF could scale to 1000s of nodes.